\documentclass[a4paper,12pt]{article}
\usepackage[british]{babel}
\usepackage{amsfonts}
\usepackage{amsfonts}
\usepackage{amsmath}
\usepackage{amssymb}
\usepackage{graphicx}
\begin{document}

\begin{center}
\baselineskip=1.5
\normalbaselineskip{\Large The role of ergodicity and mixing in the central limit theorem for Casati-Prosen triangle map variables}

{\large S\'{\i}lvio M. Duarte Queir\'{o}s}\footnote{%
email address: Silvio.Queiros@unilever.com, sdqueiro@googlemail.com}

\baselineskip=1.0 \normalbaselineskip

\textit{Unilever R\&D Port Sunlight, Quarry Road East, Wirral, CH63 3JW UK \\%
[0pt]
}

\baselineskip=1.0 \normalbaselineskip

{\small (10th December 2008)}
\end{center}

\baselineskip=1.0 \normalbaselineskip

\subsection*{\protect\bigskip Abstract}

In this manuscript we analyse the behaviour of the probability density
function of the sum of $N$ deterministic variables generated from the
triangle map of Casati-Prosen. For the case in which the map is both ergodic
and mixing the resulting probability density function quickly concurs with
the Normal distribution. When these properties are modified
the resulting probability density functions
are described by power-laws. Moreover, contrarily to what it would be
expected, as the number of added variables $N$ increases the distance to
Gaussian distribution increases. This behaviour goes against standard
central limit theorem. By extrapolation of our finite size results we
preview that in the limit of $N$ going to infinity the distribution has the
same asymptotic decay as a Lorenztian (or a $q=2$-Gaussian).

\section{Introduction}

The central limit theory has been a subject of studied within the natural
sciences for many generations. We might even state that CLT originated in
1713 with Bernoulli's weak law of large numbers \cite{bernoulli}. After
Bernoulli, de Moivre \cite{moivre}, Laplace \cite{laplace}, and Gauss,
amongst others, made crucial contributions to the establishment of the
Normal probability density function (PDF) as the stable distribution when
one considers the sum of independent and identically distributed random
variables with finite standard deviation. The stability of the Normal
distribution, the central limit theorem (CLT), was just formally established
by the Russian mathematician Lyapunov 188 years after Bernoulli \cite%
{lyapunov}. Afterwards, L\'{e}vy and Gnedenko \cite{levy,gnedenko}
generalised the CLT to account for independent and identically distributed
random variables but with infinite standard deviations, followed by broader
generalisations which include dependency between variables \cite{araujo,
hall}. With the advent of computation in the 1970s, chaos theory and
non-linear phenomena achieved huge progress. It was then possible to verify
the existence of a central limit theory for deterministic variables as well~%
\cite{beck,kaminska}. More recently, central limit theory has been the focus
of renewed interest within statistical mechanics mostly because of the
endeavour to establish the optimising PDF of non-additive entropy $S_{q}$~%
\cite{ct88} as the stable distribution for the sum of random or
deterministic variables for some special kind of correlation~\cite%
{q-clt-gauss}, on the edge of chaos~\cite{ugur,guiomar}, or even for systems
in a metastable state \cite{hmf}.

In the sequel of this manuscript we communicate results on numerical
investigations of the distribution of deterministic variables which arise
from the sum of variables generated from the triangle map introduced by
Casati and Prosen~\cite{casati}. Our analysis is performed at two different
regimes: a first illustrative one, for which the system is both \textit{%
ergodic} and \textit{mixing}, and a second in which the system is \textit{%
weakly ergodic} and apparently \textit{weakly mixing \footnote{%
By \textit{strongly ergodic} we mean systems for which the exploration rate $%
r\left( t\right) =n\left( t\right) /N$ [($n\left( t\right) $ is the number
of cells visited by a orbit until a discrete time $t$ and $N$ is the total
number of cells $\left( N\rightarrow \infty \right) $], averaged over
several initial conditions, presents the same functional form as the
exploration rate of a random model $r_{RM}\left( t\right) =1-\exp \left[ -t/N%
\right] $. Concomitantly, by \textit{strongly mixing} we refer to systems
whose correlation function, $C\left( t\right) $ decays much faster than $%
t^{-\mu }$ with $\mu \geq 1$. Weakly ergodic/mixing system do not follow
each of corresponding the features \cite{katok}.}}. In the former case, the
convergence towards the Gaussian distribution is clear and easily explained
according to the standard CLT. In the latter case, we detect an anomalous
behaviour characterised by non-Gaussian distributions which become more
distant from the Gaussian distribution as the number of variables added
increases. If we take into account the behaviour shown for phase occupancy
observables in other conservative dynamical systems, this result is against
all odds and it emphasises the very particular properties of this dynamical
system. The remaining of the manuscript is organised as follows: in Sec.~\ref%
{intro} we introduce the triangular map and some of its properties, and in
Sec.~\ref{results} we present our numerical results. Some conclusions and
remarks are made in Sec.~\ref{remarks}.

\section{The triangle map\label{intro}}

The triangle map, $z_{t+1}=T\left( z_{n}\right) $ introduced by Casati and
Prosen~\cite{casati-3,casati}, corresponds to a discrete transformation on a
torus $z=\left( x,y\right) \in \lbrack -1,1)\times \lbrack -1,1)$ with
symmetrical coordinates,%
\begin{equation}
\left\{
\begin{array}{cccc}
x_{n+1} & = & x_{n}+y_{n+1} & \left( \mathrm{mod}^{\ast }\,2\right) \\
y_{n+1} & = & y_{n}+\alpha \,\mathrm{sign\,}x_{n}+\beta & \left( \mathrm{mod}%
^{\ast }\,2\right)%
\end{array}%
\right. ,  \label{mapa}
\end{equation}%
where $\mathrm{sign\,}x=\pm 1$, and $\alpha $ and $\beta $ are real
parameters of the map. Function $\left( \mathrm{mod}^{\ast }\,2\right) $
represents a modified definition of $\mathrm{mod}\,2$ in the interval
between $-1$ and $+1$. This map has emerged from studies on the
compatibility between linear dynamical instability and the exponential decay
of Poincar\'{e} recurrences. Map (\ref{mapa}) is parabolic and area
preserving. For the Jacobian matrix we have, $\det J=1$, and its trace, $%
\mathrm{Tr}\,J=2$.

Concerning the relevance of parameters $\alpha $ and $\beta $, it is known
that when both of the parameters are irrational numbers, the map is ergodic
\cite{casati}. Moreover, it attains the ergodicity property, \textit{i.e.},
averages over time equal averages over samples, very rapidly. For such a set
of values the map is also \textit{mixing} \cite{casati-3} in the sense that
it has a \textit{continuous spectral density} besides the property previously defined.
Evaluating Poincar\'{e} recurrences, it has been found that the probability of an orbit to stay
outside a specific subset of the torus for a time longer than $t$ behaves as
$\exp \left[ -\mu \,t\right] $, with $\mu $ being very close to the Lesbegue
measure of that subset. This fact is in accordance with a completely
stochastic dynamics. The exponential decay leads to a linear separation of
close orbits which has been related to non-extensive statistical mechanics
formalism via a generalised Pesin-like identity that bridges a $q$%
-generalisation of Kolmogorov-Sinai entropy \cite{robledo} and a $q$%
-Lyapunov coefficient from the sensitivity to initial conditions~\cite%
{tsallis-csf}. The entropic index of this map was found to be $q=0$~\cite%
{casati-fulvio}.

When $\alpha =0$ and $\beta $ is irrational, the map is still ergodic \cite%
{ergodic}, but is never mixing \cite{mixing}. For the case $\beta =0$ two
situations might occur \cite{casati}. If $\alpha $ is a rational number,
then the dynamics are pseudo-integrable and confined, whereas for $\alpha $
irrational, the dynamics are found to be weakly ergodic, with the number of $%
y_{n}$ taken by a single orbit increasing as $\ln T$ ($0\leq n<T$).
Furthermore, the ultra-slow apparent decay of the correlation function
measured for this condition does not provide sufficient evidence of mixing.
In recent years, some analytical attempts in order to characterise the map
Eq.~(\ref{mapa}) have been made~\cite{slovenia}. Despite this, it has not
been possible to prove or reject the mixing property for the Casati-Prosen
map although strong indication of a non-mixing property prevails.

\section{Results\label{results}}

In this section, we present the numerical results of our study. Namely, we
have considered variables $X_{N}$ and $Y_{N}$ that are obtained from the
addition of $x$ and $y$ variables of map (\ref{mapa}),%
\begin{equation}
X_{N}=\sum_{i=1}^{N}x_{i},
\end{equation}%
\begin{equation}
Y_{N}=\sum_{i=1}^{N}y_{i},
\end{equation}%
that we analyse after detrending and rescaling, $u_{N}^{\prime }=\left(
u_{N}-\left\langle u_{N}\right\rangle \right) /\sigma _{N}$ ($u$ renders
either $X$ or $Y$ and $\sigma _{N}$ is the standard deviation, $\sigma
_{N}^{2}=\frac{1}{N}\sum_{i=1}^{N}u_{i}^{2}$). In our survey we have
neglected the case $\alpha =0$ since it destroys the dependence of $y$ on $x$
and we have focussed on the following situations,
\begin{equation}
\mathrm{case\ I:}\left\{
\begin{array}{c}
\alpha =\frac{1}{2}\left[ \frac{1}{2}\left( \sqrt{5}-1\right) -e^{-1}\right]
, \\
\beta =\frac{1}{2}\left[ \frac{1}{2}\left( \sqrt{5}-1\right) +e^{-1}\right]%
\end{array}%
\right\} ,
\end{equation}%
which corresponds to the ergodic and mixing case studied in Refs. \cite%
{casati,casati-fulvio} and
\begin{equation}
\mathrm{case\ II:}\left\{ \alpha =\pi ^{-\frac{1}{2}},\beta =0\right\} ,
\end{equation}%
where the map is weakly ergodic. For each analysis, we have randomly placed
a set of initial conditions $\mathcal{I}$ (typically $10^{7}$ elements)
within the torus $z$ and we have let the map run. The probability density
functions $P\left( X_{N}\right) $ and $P\left( Y_{N}\right) $ are then
obtained from these $\mathcal{I}$ initial conditions. Our numerical
calculations have been performed using the \textsc{Mathematica\texttrademark
} kernel which assures a symbolic computational procedure. Although
analytical considerations in case~I are in principle possible, we have
skipped them because we have used case~I as a benchmark of the peculiar
behaviour of case~II that we are going to present. It is also worth stating
that, for case~I, where analytical considerations have been made (or can be
made), conditions of strong mixing and (semi)conjugation to a Bernoulli
shift are mandatory~\cite{beck}. These conditions are not verified in our
case~II.

\begin{figure}[tbh]
\begin{center}
\includegraphics[width=0.85\columnwidth,angle=0]{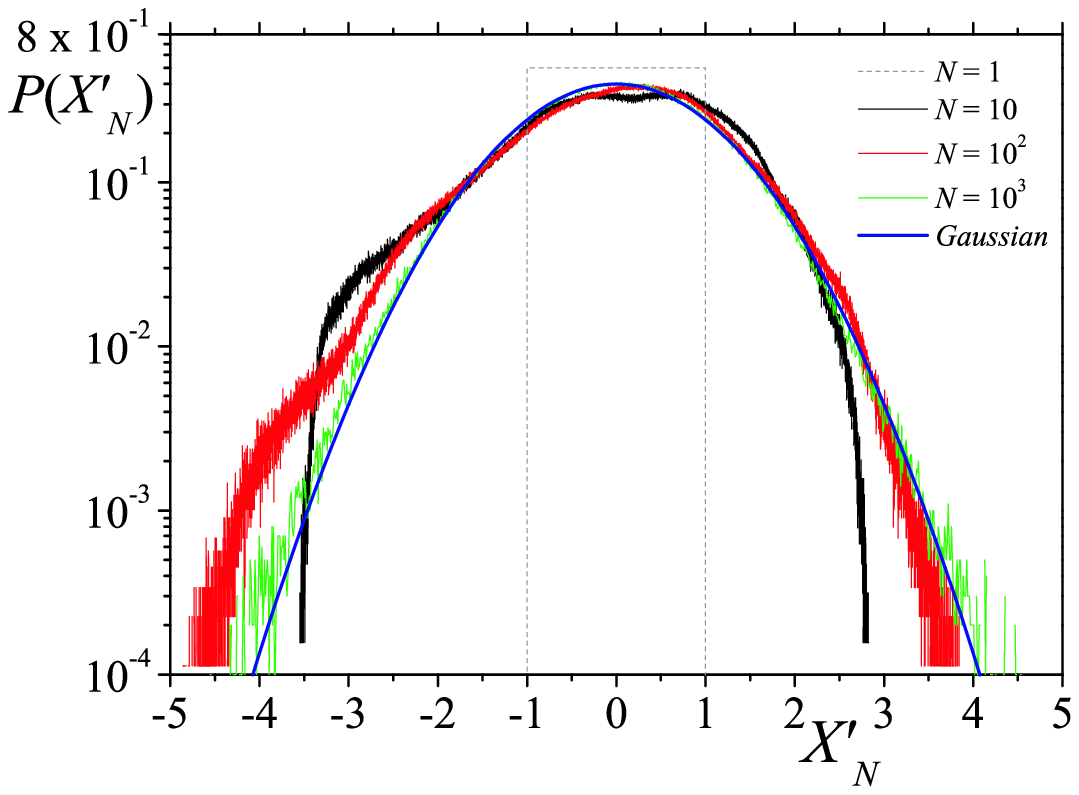} %
\includegraphics[width=0.85\columnwidth,angle=0]{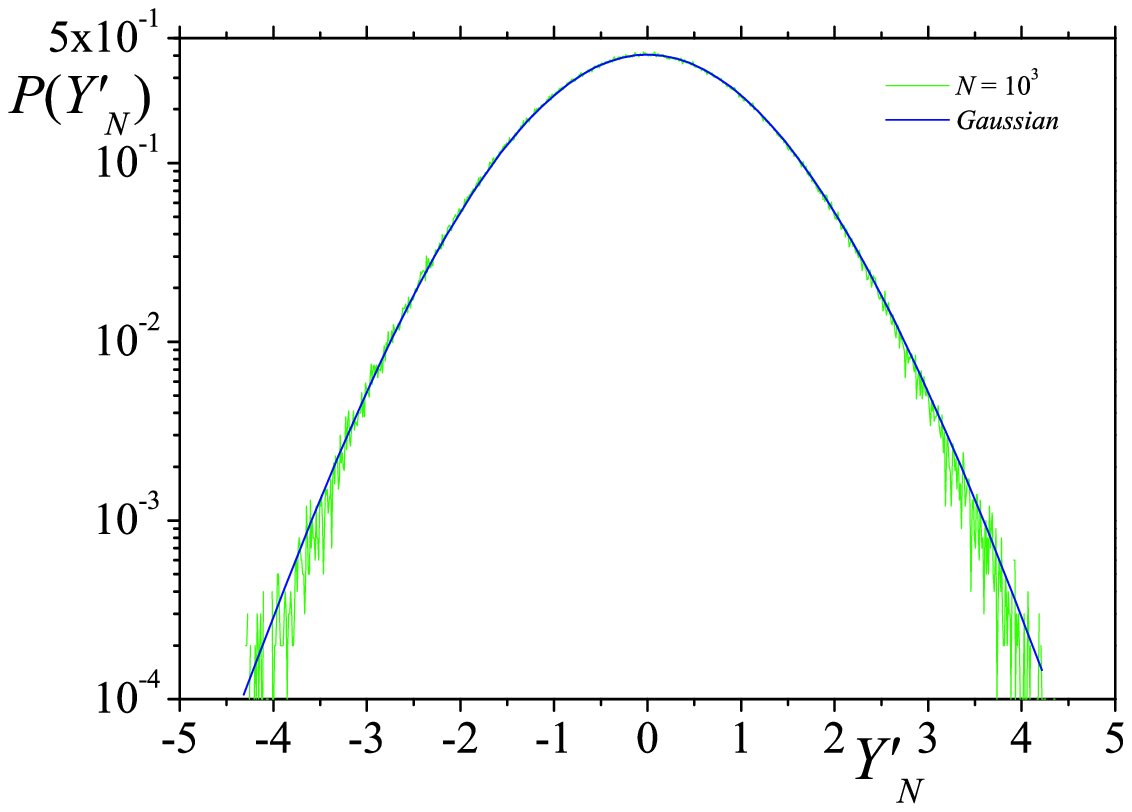}
\end{center}
\caption{Upper panel: Probability density function $P\left( X_{N}^{\prime
}\right) $ $vs$ $X_{N}^{\prime }$ for scaled variables $X_{N}^{\prime
}\equiv X_{N}-\left\langle X_{N}\right\rangle /\protect\sigma _{N}$,
obtained in case~I where $\left\langle X_{N}\right\rangle $ represents the
average of $X_{N}$ [on log-linear scale]. Lower panel: The same as the upper
panel, but for variable $Y_{N}$. In both panels the line labelled \textit{%
Gaussian} corresponds to Eq. (\protect\ref{gauss}) with $\protect\sigma %
_{N}=1$.}
\label{fig-pdf-casei}
\end{figure}

\begin{figure}[tbh]
\begin{center}
\includegraphics[width=0.85\columnwidth,angle=0]{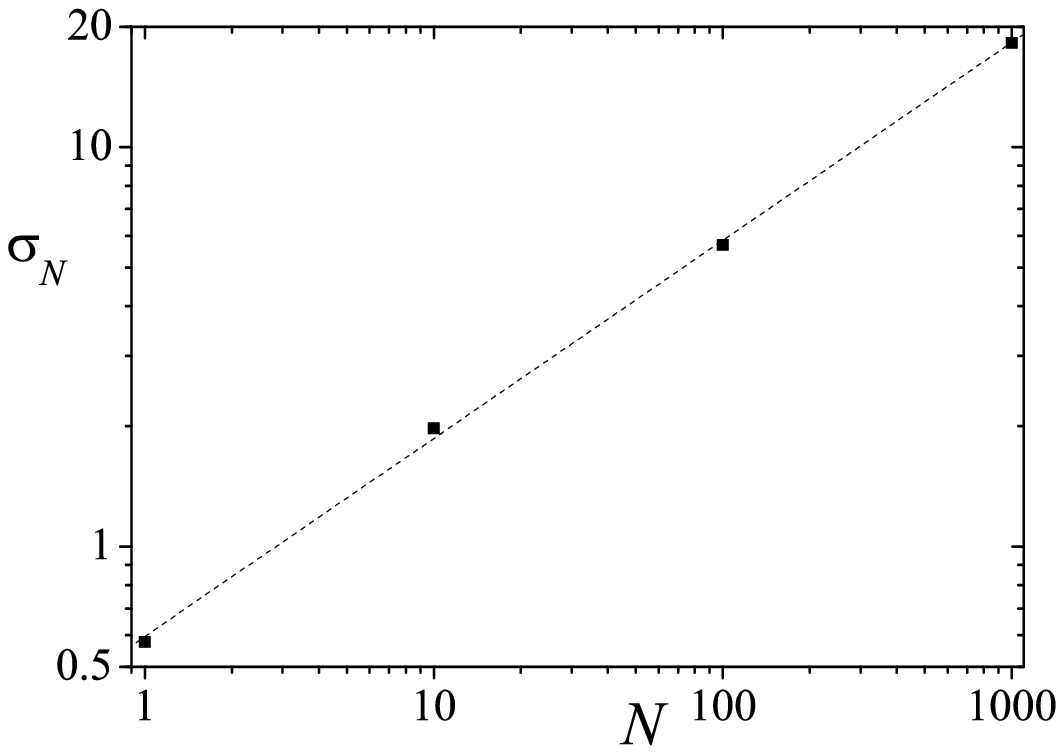}
\end{center}
\caption{Standard deviation, $\protect\sigma _{N}$, of $X_{N}$ vs. $N$ for
case~I [on $\log $-$\log $ scale]. The fitted straight line has a slope of $%
0.49\pm 0.01$.}
\label{sd-casei}
\end{figure}

For case~I, as expected from ergodic and mixing properties of map (\ref{mapa}%
), both $X_{N}$ and $Y_{N}$ (detrended) approach the Normal distribution
\cite{kaminska},%
\begin{equation}
G\left( u_{N}\right) =\frac{1}{\sqrt{2\pi \sigma _{N}^{2}}}\exp \left[ -%
\frac{1}{2\sigma _{N}^{2}}u_{N}^{2}\right] ,  \label{gauss}
\end{equation}%
as $N$ goes to infinity, see Fig. \ref{fig-pdf-casei}. Moreover, as it
occurs for the Lyapunov CLT, $\sigma _{N}$ follows the scaling relation,
\begin{equation}
\sigma _{N}=\sqrt{N}\sigma _{1},  \label{sigmascaling}
\end{equation}%
with $\sigma _{1}=\frac{1}{\sqrt{3}}$ as in Fig.~\ref{sd-casei}. In Ref.~%
\cite{casati} a similar kind of analysis has been made by considering a
cylinder $y\in \left( -\infty ,\infty \right) $
\begin{equation}
y_{n}=y_{0}+\beta \,n+\alpha \,p_{n},  \label{torus}
\end{equation}%
with $p_{n}\in \mathbb{Z}$.

We have also verified a skew in our PDFs, for small $N$, which are not
visible in the PDFs of Ref.~\cite{casati}, but might be comprehended
according to analytical work made on other types of maps~\cite{beck,ugur}.
Explicitly, skewed distributions have \textit{analytically} been found when
studying the same problem using the dissipative, fully chaotic and strongly
mixing map $x_{t+1}=1-2\,x_{t}^{2}$~\cite{ugur}.

\medskip

\begin{figure}[tbh]
\begin{center}
\includegraphics[width=0.85\columnwidth,angle=0]{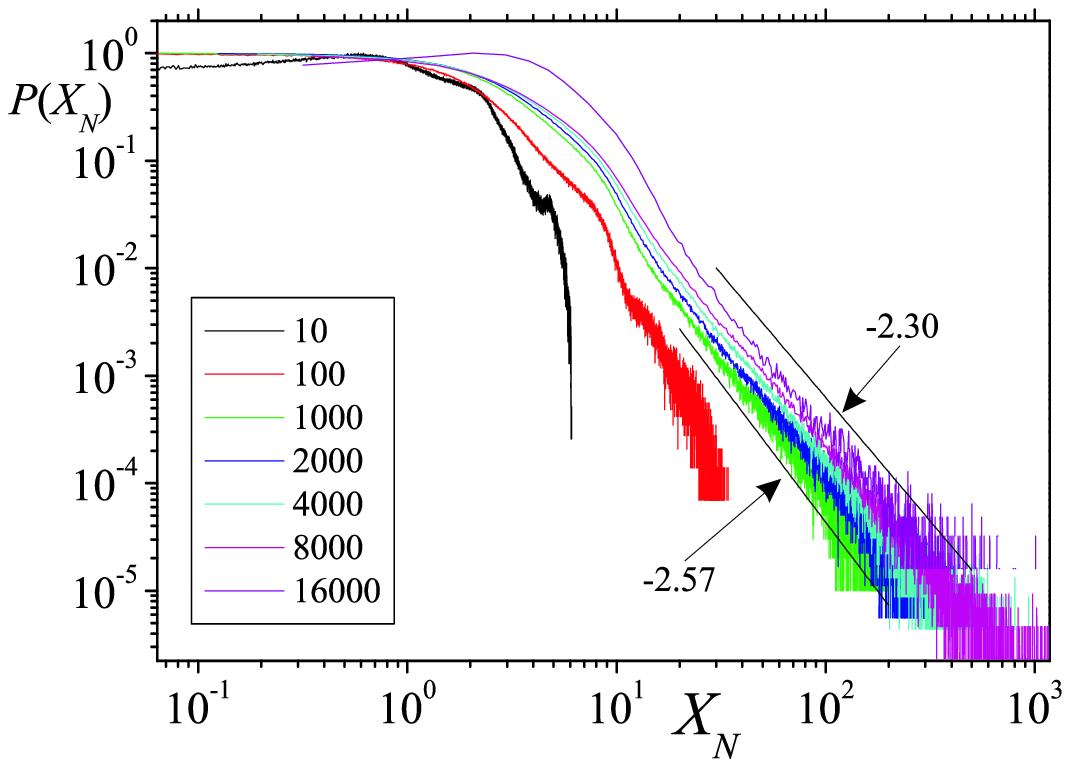}
\end{center}
\caption{Probability density function $P^{\prime }\left( X_{N}\right) =\frac{%
P\left( X_{N}\right) }{P\left( X_{N}\right) _{\max }}$ \textit{vs} $X_{N}$
[in $\log $-$\log $ scale]. The number of initial conditions (points used to
construct PDFs) is $10^{7}$ except for $N=8000$ ($7.5\times 10^{6}$ points)
and $N=16000$ ($7\times 10^{5}$ points). The power-law decay is evident for
large $X_{N}$. The two straight lines have slopes $-\left( \protect\eta %
_{+}+1\right) $ with $\protect\eta $ the exponent for $N=1000$ and $N=16000$
(see Table \protect\ref{table}).}
\label{fig-pdf-caseii}
\end{figure}

A completely different behaviour is found when case~II is analysed. For this
case, we have concentrated on $X_{N}$, although for $Y_{N}$ we have obtained
the same qualitative results. Instead of distributions reminiscent of a
Normal distribution, we have numerically observed probability distributions
which are well described by,
\begin{equation}
P\left( u\right) \sim \left\vert u\right\vert ^{-\eta -1}\qquad \left( 1\ll
u\leq N\right) ,  \label{pdfcaseii}
\end{equation}%
with $\eta <2$ for every value of $N$ analysed, see Tab.~\ref{table} and
Figs. \ref{fig-pdf-caseii}\ and \ref{fig-table}. The method applied to
determine $\eta $ has been the Meersch\"{a}rt-Scheffler estimator (see
Appendix)~\cite{estimador}. The subsequent application of the Hill estimator~%
\cite{hill} has given concordant results.

\begin{figure}[tbh]
\begin{center}
\includegraphics[width=0.85\columnwidth,angle=0]{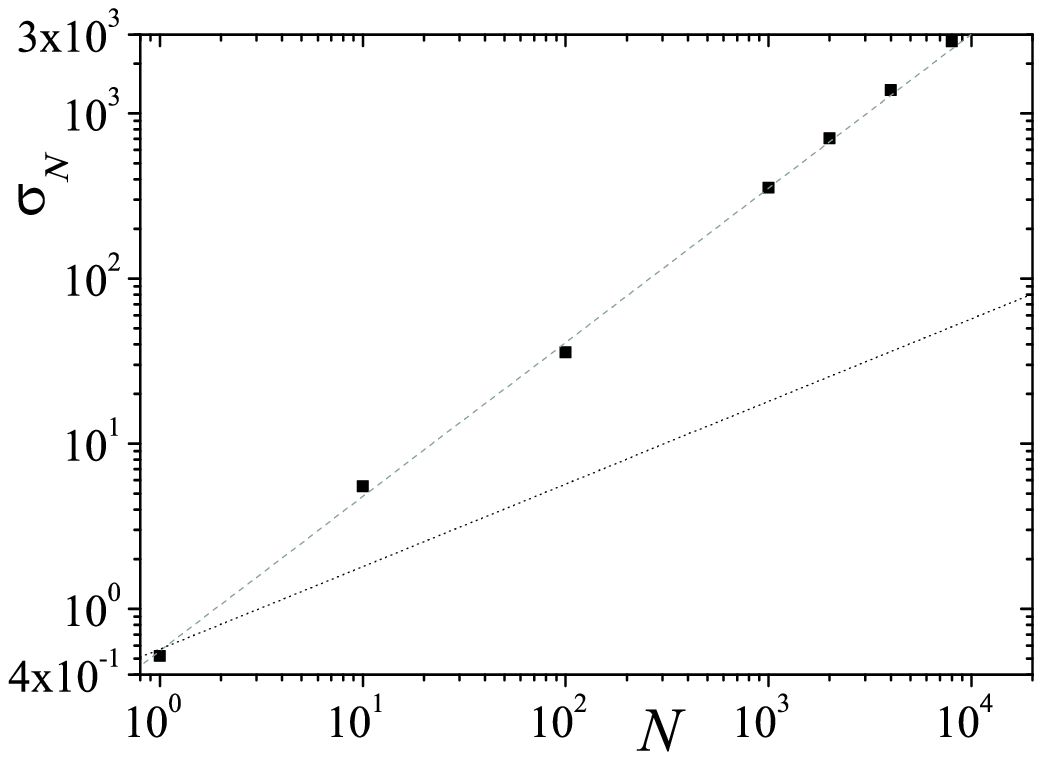}
\end{center}
\caption{Standard deviation $\protect\sigma _{N}$ \textit{vs} $N$ of $X_{N}$
obtained from map (\protect\ref{mapa}) [on $\log $-$\log $ scale]. The
dotted line corresponds to Eq.~\protect\ref{sigmascaling} and the grey
dashed line has slope $0.94 \pm 0.02$ corresponding to the best linear fit
on this particular scale.}
\label{sd-caseii}
\end{figure}

The upper bound of $\eta $ we have found ($\eta =1.80$) imposes that the
standard deviation would diverge if the variable $X$($Y$) was defined over
the whole interval of real numbers. Since we are treating cases for which $N$
is finite, the support of the resulting PDFs is compact and defined between $%
-N$ and $N$ for $X_{N}$ and $Y_{N}$. This obviously leads to a finite
standard deviation, $\sigma _{N}$, for the cases we have studied. We have
verified that $\sigma _{N}$ does not follow the scaling relation Eq.~(\ref%
{sigmascaling}), see Fig.~\ref{sd-caseii}. Instead, a power-law dependence
can approximately be given with an exponent close to $0.94\pm 0.02$. In
addition, we have observed that the shape of the distribution $P\left(
X_{N}\right) $ has strong similarity with the $\alpha $-stable L\'{e}vy
distribution
\begin{equation}
\mathcal{L}_{\alpha }\left( X_{N}\right) =\frac{1}{2\,\pi }\int_{-\infty
}^{+\infty }\exp \left[ -i\,k\,X_{N}-a\left\vert k\right\vert ^{\alpha }%
\right] dk,  \label{levy}
\end{equation}%
($\left( 0<\alpha <2\right) $)when plotted in a log-log scale, namely the
emergence of an inflexion point before the straight line segment~\cite%
{tsallis-queiros}. We must emphasise that this similarity \textit{by no
means implies a possible application of L\'{e}vy-Gnedenko generalisation of
the CLT} which states that, the probability density function of the sum of $%
N $ variables, each one associated with the same distribution Eq.~(\ref%
{pdfcaseii}) ($0<\eta <2$ and $u\in \Re $) converges in the limit as $N$
goes to infinity to a $\alpha $-stable L\'{e}vy distribution with $\alpha
=\eta $. It is easy to verify that the results we report in this manuscript
do not follow this theorem because; \textit{i)} $x_{t}$ and $y_{t}$
variables are boxed up within intervals from $-1$ to $+1$ and hence their
standard deviations are always finite, \textit{ii)} due to the weak chaotic
properties, variables $x_{t}$ and $y_{t}$ are not idenpendent at all
instants $t$.

\begin{figure}[tbh]
\begin{center}
\includegraphics[width=0.85\columnwidth,angle=0]{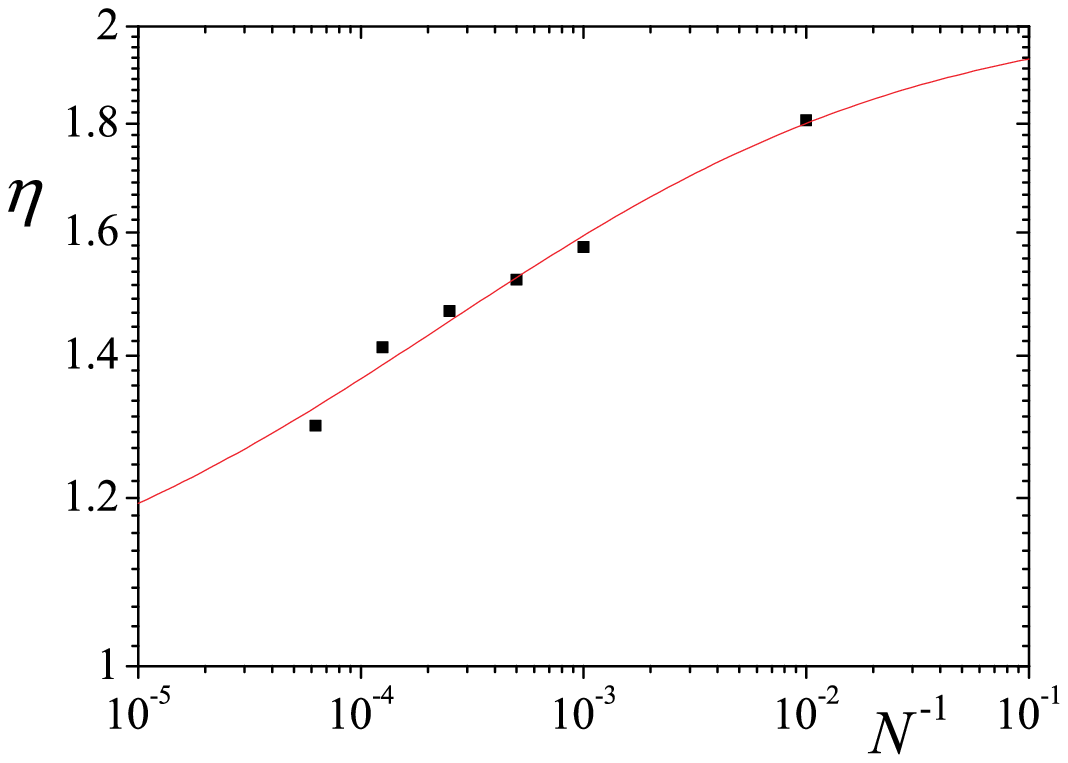}
\end{center}
\caption{Values of the exponent for positive side, $\protect\eta _{+}$, vs. $%
N^{-1}$ [on log-log scale]. The line represents a numerical adjustment for
Eq. (\protect\ref{extrapolate}) with $\protect\eta _{\infty }=1.02$, $c=0.049
$, and $\protect\delta =0.40$ ($\protect\chi ^{2}=6.9\times 10^{-4}$ and $%
R^{2}=0.986$).}
\label{fig-table}
\end{figure}

\begin{figure}[tbh]
\begin{center}
\includegraphics[width=0.85\columnwidth,angle=0]{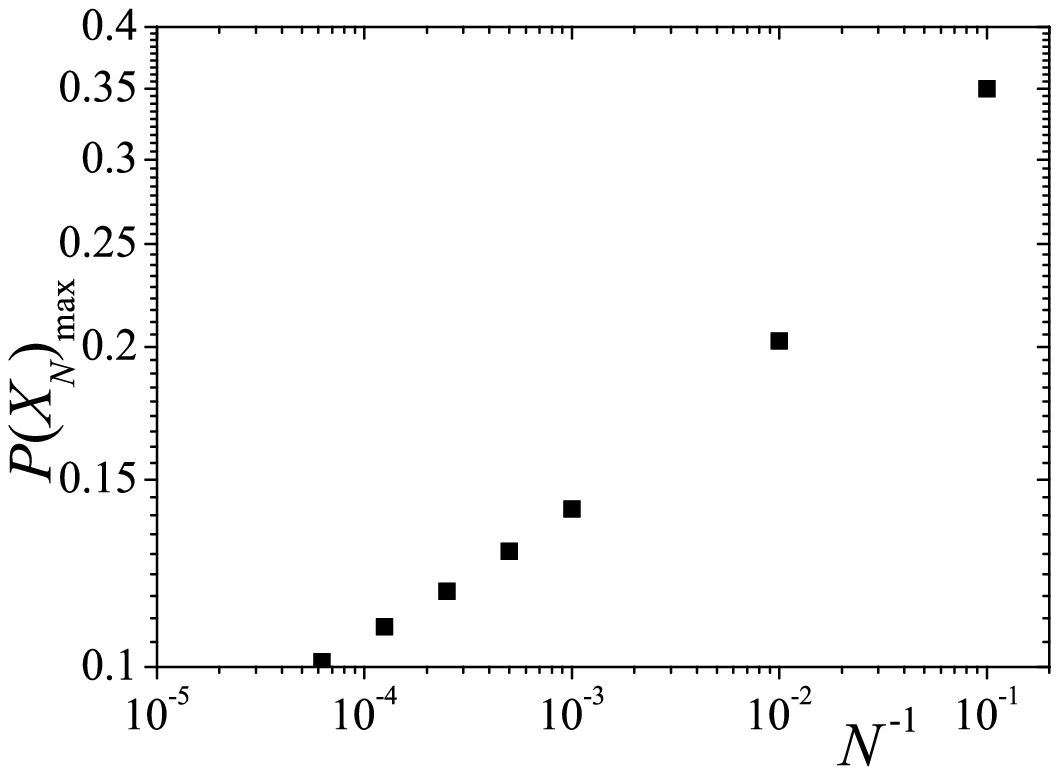}
\end{center}
\caption{Maximum value $P\left( X_{N}\right) _{\max }$ vs. $N^{-1}$ according to
the values of Table \protect\ref{table} [on log-log scale].}
\label{fig-pxnmax}
\end{figure}

Trying to infer about the scaling behaviour of $P\left(X_{N}\right) _{\max }$
with $N$, a clear-cut power-law behaviour could not be found as it is
visible in Fig.~\ref{fig-pxnmax}.

In the absence of clear power-law behaviour and using the fact that $\eta $
decreases as $N$ increases, we have tried to extrapolate a value of $\eta
\left( N\rightarrow \infty \right) $. To this end, inspired by finite-size
scaling relations of critical behaviour~\cite{herrmann}, we have used the
following ansatz,
\begin{equation}
\eta \left( N\right) =\eta _{\infty }\left( 1+\frac{1}{1+cN^{\delta }}%
\right) .  \label{extrapolate}
\end{equation}%
From this, we have obtained $\eta _{\infty }=1.02\pm 0.06$ (see Fig. \ref%
{fig-table}) very close (within error margins) to the exponent of the
Lorentzian distribution, $\mathcal{L}_{1}\left( X_{N}\right) $.

\begin{table}[tbp]
\caption{Values of the maximum value of $P\left( X_{N}\right) _{\max }$, and
$\protect\eta$ parameters characterising PDF Eq.~(\protect\ref{pdfcaseii})
for positive, $\protect\eta _{+}$, and negative, $\protect\eta _{-}$,
branches for each value of $N$ calculated.}
\label{table}
\begin{center}
\begin{tabular}{cccc}
\hline
$N$ & $P\left( X_{N}\right) _{\max }$ & $\eta _{+}$ & $\eta _{-}$ \\
\hline\hline
$10$ & $0.350$ &  &  \\
$10^{2}$ & $0.203$ & $1.80$ & $1.81$ \\
$10^{3}$ & $0.141$ & $1.57$ & $1.57$ \\
$2\times 10^{3}$ & $0.128$ & $1.52$ & $1.52$ \\
$4\times 10^{3}$ & $0.118$ & $1.47$ & $1.47$ \\
$8\times 10^{3}$ & $0.109$ & $1.41$ & $1.41$ \\
$1.6\times 10^{4}$ & $0.101$ & $1.30$ & $1.29$ \\ \hline
\end{tabular}%
\end{center}
\end{table}

\section{Final remarks\label{remarks}}

In this manuscript we have presented a numerical experiment on the addition
of deterministic variables generated by a conservative map, the triangle map
of Casati and Prosen~\cite{casati}. The study has been performed in two
different regimes, case~I and case~II, by iterating the map from a set of
initial conditions which have uniformly been placed within interval $[-1,1)$%
. In case~I, for which the map is ergodic and mixing, the outcoming stable
PDF is the Normal distribution for both $X_{N}$ and $Y_{N}$, in perfect
accordance with standard theory. In case~II, for which the map is weakly
ergodic for sure and with apparently no mixing, we have obtained PDFs which
are well described by power-laws for large values of the variable. Moreover,
the parameter characterising the PDF is smaller than $3$ and it presents a
decreasing trend as the number of variables $N$ are augmented \footnote{%
An exponent equal to $3$ corresponds to the lower bound for finite second
order moment of distributions defined between $-\infty $ and $\infty $.}.
Our results are to some extent surprising seeing that, notwithstanding the
map is weakly ergodic, it fills the phase space as time evolves.
Accordingly, it would be expected that Gaussian behaviour would be
approached (as in case~I) as $N$ increases and not the opposite as we have
reported here. In such a scenario of weak chaoticity, the map was expected
to present some anomalous (quasi-steady) behaviour before total occupancy of
the phase space took place which would imply a crossover to the Normal
distribution. This last description is analogous to what has been observed
in the fractal dimension, which tends to increase towards the Euclidean
dimension, for the case of low-dimensional sympletic maps (see details in
Ref.~\cite{fulvio-edgardo}). However, for the case we have studied, the
observed increasing departure from the Gaussian distribution with increasing
$N$ points to another direction\footnote{$N$ can also work as a measure of
time.}. Last of all, from the ansatz~\ref{extrapolate}, we have extrapolated
the value of $\eta $ when $N\rightarrow \infty $, which has shown to be $%
\eta _{\infty }\approx 1$, \textit{i.e.}, the same decay as the Lorentz
distribution, which corresponds to a $q$-Gaussian with $q=2$ in the
non-extensive formalism. In defiance of the different nature of both
systems, our result for the Casati-Prosen conservative system has provided a
similar qualitative result as obtained in Refs.~\cite{ugur} for the logistic
(dissipative) map. In other words, departing from variables with a finite
standard deviation that evolve according with a dynamical system in which
strongly ergodic and strongly mixing features are not verified, we have been
able to define a new variable whose limit distribution (spanning the whole
domain of real numbers) has a different attractor that for the Gaussian.
Moreover, in our results the mixing property looks to be a crucial element.
It is our expectation that, along with results for dissipative dynamical
systems, this work should suggest further studies on the introduction of an
analytical framework for cases where the Bernouilli shift is not verified.

\subsection*{Acknowledgements}

SMDQ would like to thank Prof C. Tsallis for several discussions over
several aspects of central limit theory, G.~Ruiz for conversations about the
application of CLT to deterministic variables, and T. Prosen for his
comments on the properties of Eq.~(\ref{mapa}) and for having provided his
work~\cite{slovenia} before public disclosure. An acknowledgment is also
addressed to Prof T. Cox for the thorough reading and commenting of this
manuscript. 
The work herein presented has benefited from financial support by FCT/MCTES
(Portuguese agency) and Marie Curie Fellowship Programme (European Union),
and infrastructural support from PRONEX/CNPq (Brazilian agency).

\section*{Appendix}

The method introduced by Meerschaert and Scheffler \cite{estimador} is based
on the asymptotic limit of the sum of the variables of dataset $\left\{
X_{N}\right\} $ under scrutiny. For heavy tail data these asymptotics depend
only on the tail index of the probability density function, and not on the
exact form of the distribution. Hence, if $\mathcal{I}$ elements of a
dataset are identically and (in)dependently distributed, and in addition its
probability density function presents an asymptotic behaviour,%
\begin{equation*}
P\left( X_{N}\right) \sim \left\vert X_{N}\right\vert ^{-\eta -1},\qquad
\left( \left\vert X_{N}\right\vert \rightarrow \infty \right) ,
\end{equation*}%
it can be proved (Theorem 1 in Ref. \cite{estimador}) that,

\begin{equation}
\frac{1}{\eta }=\frac{\ln _{+}\left[ \Sigma _{i=1}^{\mathcal{I}}\left(
X_{N,i}-\left\langle X\right\rangle \right) ^{2}\right] }{2\ln \mathcal{I}},
\tag{A1}
\end{equation}%
where $\left\langle X\right\rangle $ is the simple average $\left\langle
X\right\rangle =\mathcal{I}^{-1}\Sigma _{i=1}^{\mathcal{I}}X_{N,i}$ and $\ln
_{+}\left[ x\right] \equiv \max \left\{ \ln x,0\right\} $.

\end{document}